Growth and superconducting properties of F-substituted ROBiS$_2$ (R = La, Ce, Nd) single crystals

Masanori Nagao[a,b,*], Akira Miura[a], Satoshi Demura[b], Keita Deguchi[b], Satoshi Watauchi[a], Takahiro Takei[a], Yoshihiko Takano[b], Nobuhiro Kumada[a], and Isao Tanaka[a]

[a]*University of Yamanashi, 7-32 Miyamae, Kofu, Yamanashi 400-8511, Japan*

[b]*National Institute for Materials Science, 1-2-1 Sengen, Tsukuba, Ibaraki 305-0047, Japan*

*Corresponding Author

Masanori Nagao

Postal address: University of Yamanashi, Center for Crystal Science and Technology

Miyamae 7-32, Kofu 400-8511, Japan

Telephone number: (+81)55-220-8610

Fax number: (+81)55-254-3035

E-mail address: mnagao@yamanashi.ac.jp

**Abstract**

F-substituted ROBiS$_2$ (R = La, Ce, Nd) superconducting single crystals with different F concentration were grown successfully using CsCl/KCl flux. All the obtained single crystals had a plate-like shape with a well-developed *ab*-plane of 1-2 mm in size. The flux components of Cs, K, and Cl were not detected in the obtained single crystals by electron probe microanalysis. The grown single crystals of F-substituted LaOBiS$_2$ and CeOBiS$_2$ showed superconducting at about 3 K while the $T_c$ of the F-substituted NdOBiS$_2$ exhibited approximately 5 K. The superconducting anisotropy of the single crystals of F-substituted LaOBiS$_2$ and NdOBiS$_2$ was estimated to be 30-45 according to the effective mass model whereas those values were 13-21 for the F-substituted CeOBiS$_2$ single crystals. The F-substituted CeOBiS$_2$ single crystals exhibited magnetic order at about 7 K that apparently coexisted with superconductivity below around 3 K.

**Main text**

**1. Introduction**

The discovery of the BiS$_2$-based superconductors, Bi$_4$O$_4$S$_3$ [1] and RO$_{1-z}$F$_z$BiS$_2$ (R = La, Ce, Pr, Nd, Yb) [2-6], has attracted a great interest as new superconductors. The substitution of F for O in these materials are important for the appearance of superconductivity by inducing carriers in superconducting layers. However, the investigation of these superconductors has faced problems of the lack of single crystals. Very recently, we have reported the growth of single crystal of NdO$_{0.7}$F$_{0.3}$BiS$_2$ by CsCl/KCl flux in vacuum and its intrinsic transport properties [7]. This successful growth of the single crystal using CsCl/KCl flux motivates us to grow other BiS$_2$-based single crystals. It is highly desirable to grow the single crystals with various compositions in order to investigate the effect of different rare earth elements and F contents in F-substituted ROBiS$_2$ on their intrinsic properties.

In this paper, we grew F-substituted ROBiS$_2$ (R = La, Ce, Nd) single crystals with different F concentration by using CsCl/KCl flux. The composition and transport properties were examined to understand intrinsic properties of these materials, and we discussed the effect of rare earth elements and F substitution.

**2. Experimental**

Single crystals of F-substituted ROBiS$_2$ (R:La,Ce,Nd) were grown by a high-temperature flux method in a vacuumed quartz tube. The raw materials were R$_2$S$_3$, Bi, Bi$_2$S$_3$, Bi$_2$O$_3$, BiF$_3$, CsCl, and KCl. The synthesis and characterization procedures followed the previous literature except the amount of CsCl/KCl flux [7]. The raw materials were weighed with a nominal composition of RO$_{1-x}$F$_x$BiS$_2$ ($x$ = 0-1.0). The mixture of the raw materials (0.8 g) and CsCl/KCl flux (5.0 g) were mixed using a mortar, and sealed in a quartz tube in vacuum. The molar ratio of the CsCl/KCl flux was CsCl:KCl = 5:3. The mixed powder was heated at 800 °C for 10 h, cooled slowly to 600 °C at a rate of 1 °C/h, and then furnace-cooled down to room temperature. The quartz tube was opened in air atmosphere, and the flux was solved into distilled water in the quartz tube. The product was filtered, and washed with distilled water. The crystal structure and composition of the single crystals were evaluated by X-ray diffraction (XRD) analysis using CuK$_\alpha$ radiation, scanning electron microscopy (SEM), and electron probe microanalysis (EPMA). The transport properties of the single crystals were measured by the standard four-probe method with constant current mode using a Physical Property Measurement System (Quantum Design; PPMS DynaCool). The electrical terminals were made by silver paste. We measured the angular ($\theta$) dependence of resistivity ($\rho$) in the flux liquid state under various magnetic fields ($H$) and calculated the superconducting anisotropy ($\gamma_s$) using the effective mass model [8-10]. The temperature dependence of magnetization ($M$) under zero-field cooling (ZFC) and field cooling (FC) was measured by a superconducting quantum interface device (SQUID) magnetometer with an applied magnetic field of 10 Oe parallel to the *ab*-plane.

## 3. Results and discussion

Figure 1 shows a typical SEM image of the F-substituted $CeOBiS_2$ single crystal. The synthesized single crystals had a plate-like shape of 1.0-2.0 mm in size and 10-20 μm in thickness. The in-plane orientation of the grown crystals was not measured. Figure 2 shows the XRD pattern of a well-developed plane of the F-substituted $CeOBiS_2$ single crystal grown from the starting powder with a nominal composition of $CeO_{0.7}F_{0.3}BiS_2$. The peak positions of the grown single crystals agree with those of *00l* diffraction peaks of F-substituted $CeOBiS_2$ in the previous literature [3]. The presence of only *00l* diffraction peaks of the $CeOBiS_2$ structure indicates the well-developed *ab*-plane.

Table I shows the effect of the F composition in the starting materials on the F

composition of the grown single crystals, the lattice parameter of *c*-axis and superconducting anisotropy ($\gamma_s$) in the single crystals. The F composition, which is normalized by total of the F and O compositions, in the starting materials and in the single crystals are defined as *x* and *y*, respectively. The single crystals were obtained only in the range of *x* between 0.3 and 0.9. The chemical ratio of R:Bi:S in the grown crystals determined to be 1.01±0.05:0.99±0.05:2.00 by EPMA, and the ratio agreed with the stoichiometric ratio. The obtained values were normalized using S = 2.00. Nd and Bi were measured to a precision of two decimal places. No Cs, K, and Cl were detected in the grown single crystals by EPMA with a minimum sensitivity limit of 0.1 wt%. With increasing the F composition *x* in the starting materials, the F composition *y* in the single crystals increased and saturated to 0.46 (R = La), 0.66 (R = Ce) and 0.38 (R = Nd) at *x* ≥ 0.7. Although the analytical values of F composition *y* in the grown single crystals of *x* = 0.7 and 0.9 are consistent within analytical error, the physical properties of these crystals are different each other as described later. The increase of *x* in the starting materials shortened the lattice parameters but gave the almost same values at *x* ≥ 0.7. The minimum *c*-axis lattice parameter is about 13.4 Å regardless of rare earth metals. Figure 3 showed the *004* diffraction peaks of the F-substituted ROBiS$_2$ (R:La, Ce, Nd) single crystals. The peak position shifted toward high angle with increasing the value of *x*, and this result suggests the increase of *c*-axis lattice parameters as shown in Table I.

Figure 4 showed the temperature dependence of resistivities in the single crystals of F-substituted ROBiS$_2$ (R:La, Ce, Nd) at 2-10 K. The similar trends were observed in the single crystals of F-substituted LaOBiS$_2$ and CeOBiS$_2$. While these LaOBiS$_2$ and CeOBiS$_2$ crystals for *x* = 0.3 did not exhibit zero resistivity down to 2 K, those for *x* = 0.7 and 0.9 showed superconducting transition temperature ($T_c$) about 3 K. The transport properties of the single crystals for *x* = 0.3, 0.7 and 0.9 showed semiconducting behavior, but LaOBiS$_2$ for *x* = 0.9 became metallic. Only in the single crystals of CeOBiS$_2$, the increase of *x* from 0.7 to 0.9 significantly suppressed $T_c$ though the value of *y* in the crystals was almost same. The origin of this difference is unclear and further examination is needed. On the other hand, F-substituted NdOBiS$_2$ crystals for *x* = 0.3, 0.7 and 0.9 were superconductor with $T_c$ of approximately 5 K, and all the single crystals showed metallic behavior. The resistivities of normal state seems independent over *x* and *y* values. This may be attributed to the increase of surface roughness. We could observe the rough surface of these crystals to the naked eyes when we compared with the surface of F-substituted LaOBiS$_2$ and CeOBiS$_2$ single crystals. The broad transition for *x* = 0.7 might be due to the low crystallinity.

The angular ($\theta$) dependence of resistivity ($\rho$) was measured at different magnetic fields ($H$) in the flux liquid state to estimate the superconducting anisotropy ($\gamma_s$) of grown F-substituted ROBiS$_2$ single crystals, as reported in Refs. 8 and 9. The reduced field ($H_{red}$) is calculated using the following equation for the effective mass model:

$$H_{red} = H(\sin^2\theta + \gamma_s^{-2}\cos^2\theta)^{1/2} \qquad (1)$$

where $\theta$ is the angle between the *ab*-plane and the magnetic field [10]. $H_{red}$ is calculated from $H$ and $\theta$. The superconducting anisotropy ($\gamma_s$) was estimated from the best scaling for the $\rho$-$H_{red}$ relations. Figure 5 displays the angular ($\theta$) dependence of resistivity ($\rho$) at different magnetic fields ($H$ = 0.1-5.0 or 0.1-9.0 T) in the flux liquid state for the synthesized single crystals of (a) F-substituted LaOBiS$_2$ grown from the starting materials of $x$ = 0.9, $y$ = 0.46 and (b) F-substituted CeOBiS$_2$ for $x$ = 0.7, $y$ = 0.65. The $\rho$-$\theta$ curve exhibited twofold symmetry. Figure 6 shows the $\rho$-$H_{red}$ scaling obtained from the $\rho$-$\theta$ curves in Fig. 5 using Eq. (1). The scaling was performed by taking $\gamma_s$ = 45 and $\gamma_s$ = 16, shown in Fig. 6 (a) and (b), respectively. Table I shows the summary of the superconducting anisotropy of F-substituted ROBiS$_2$ single crystals with various $x$ ratios in the starting materials and $y$ ratio of the grown single crystals. While the $\gamma_s$ of R = La and Nd were estimated to be 30-45, the values of R = Ce became lower values of 13-21. Figure 7 exhibits the temperature dependence of magnetization for the obtained

F-substituted CeOBiS$_2$ single crystals for (a) $x = 0.3$, $y = 0.53$, (b) $x = 0.7$, $y = 0.65$ and (c) $x = 0.9$, $y = 0.66$ in the starting materials. The magnetic transitions ($T_m$) were only observed in the superconducting crystals for $x = 0.7$ and 0.9, and their temperature were above $T_c$. The appearance of this magnetic transition in the single crystals agrees with that in the polycrystals with similar compositions [3,6], and gives the evidence that this magnetic transition is their intrinsic property. Further characterization of these single crystals is needed for the investigation of this origin.

## 4. Conclusions

We succeeded the growth of F-substituted ROBiS$_2$ (R: La,Ce,Nd) single crystals by using CsCl/KCl flux with different rare earth metals and F substituted contents. The increase of F contents in the starting materials enhanced substituted values of F in the single crystals, but the further increase from 70 % in the starting materials gave the maximum substituted values in the single crystals. These maximum substituted values in the single crystals depended on different rare earth elements. The transport properties of the obtained single crystals were investigated. While their superconducting anisotropies ($\gamma_s$) of the single crystals of RO$_{1-y}$F$_y$BiS$_2$ (R:La,Nd $y = 0.23$-$0.46$) were estimated to be 30-45, that of R = Ce ($y = 0.53$-$0.66$) became lower values of 13-21. The F-substituted CeOBiS$_2$ single crystals exhibited magnetic order at about 7 K in addition to the superconducting transition at approximately 3 K. Further investigation of obtained single crystals is needed for the origin of these exceptional properties.

**Acknowledgments**

The authors would like to thank Drs. M. Fujioka and H. Okazaki of the National Institute for Materials Science for useful discussions.

**References**

[1]   Y. Mizuguchi, H. Fujihisa, Y. Gotoh, K. Suzuki, H. Usui, K. Kuroki, S. Demura, Y. Takano, H. Izawa, O. Miura, Phys. Rev. B **86** (2012) 220510(R).

[2]   Y. Mizuguchi, S. Demura, K. Deguchi, Y. Takano, H. Fujihisa, Y. Gotoh, H. Izawa, O. Miura, J. Phys. Soc. Jpn. **81** (2012) 114725.

[3]   J. Xing, S. Li, X. Ding, H. Yang, H.-H. Wen, Phys. Rev. B **86** (2012) 214518.

[4]   R. Jha, A. Kumar, S. Kumar Singh, V. P. S. Awana, J. Supercond. Novel Magn. **26** (2013) 499.

[5]   S. Demura, Y. Mizuguchi, K. Deguchi, H. Okazaki, H. Hara, T. Watanabe, S. J. Denholme, M. Fujioka, T. Ozaki, H. Fujihisa, Y. Gotoh, O. Miura, T. Yamaguchi, H. Takeya, Y. Takano, J. Phys. Soc. Jpn. **82** (2013) 033708.

[6]   D. Yazici, K. Huang, B. D. White, A. H. Chang, A. J. Friedman, M. B. Maple, Phil. Mag. **93** (2013) 673.

[7]   M. Nagao, S. Demura, K. Deguchi, A. Miura, S. Watauchi, T. Takei, Y. Takano, N. Kumada, I. Tanaka, J. Phys. Soc. Jpn. **82** (2013) 113701.

[8]   Y. Iye, I. Oguro, T. Tamegai, W. R. Datars, N. Motohira, K. Kitazawa, Physica C **199** (1992) 154.

[9]   H. Iwasaki, O. Taniguchi, S. Kenmochi, N. Kobayashi, Physica C **244** (1995) 71.

[10]   G. Blatter, V. B. Geshkenbein, A. I. Larkin, Phys. Rev. Lett. **68** (1992) 875.

**Figure captions**

Figure 1. Typical SEM image of F-substituted CeOBiS$_2$ single crystal.

Figure 2. XRD pattern of well-developed plane of F-substituted CeOBiS$_2$ single crystal grown from the starting powder with nominal composition of CeO$_{0.7}$F$_{0.3}$BiS$_2$.

Figure 3. (Color on the Web only) XRD pattern of *004* diffraction peaks of the F-substituted ROBiS$_2$ (R:La, Ce, Nd) single crystals.

Figure 4. (Color on the Web only) Temperature dependence of resistivities along the *ab*-plane for F-substituted (a) LaOBiS$_2$ , (b) CeOBiS$_2$, (c) NdOBiS$_2$ single crystals at 2-10 K.

Figure 5. Angular $\theta$ dependence of resistivity $\rho$ in flux liquid state at various magnetic fields (bottom to top, (a) 0.1, 0.3, 0.5, 0.7, 0.9, 1.0, 1.5, 2.0, 2.5, 3.0, 5.0, 7.0, and 9.0 T, (b) 0.1, 0.3, 0.5, 0.7, 0.9, 1.0, 3.0, and 5.0 T) for the single crystals of (a) F-substituted LaOBiS$_2$ grown from the starting materials of $x = 0.9$, $y = 0.46$ and (b) F-substituted CeOBiS$_2$ of $x = 0.7$, $y = 0.65$.

Figure 6. Data in Fig. 5 after scaling of angular $\theta$ dependence of resistivity $\rho$ at a reduced magnetic field of $H_{red} = H(\sin^2\theta + \gamma_s^{-2}\cos^2\theta)^{1/2}$. (a) F-substituted LaOBiS$_2$ of $x = 0.9$, $y = 0.46$. (b) F-substituted CeOBiS$_2$ of $x = 0.7$, $y = 0.65$.

Figure 7. Temperature dependence of magnetization under zero-field cooling (ZFC) and field cooling (FC) with an applied field of 10 Oe parallel to the *ab*-plane for the obtained single crystal using F-substituted $CeOBiS_2$ of (a) $x = 0.3$, $y = 0.53$, (b) $x = 0.7$, $y = 0.65$ and (c) $x = 0.9$, $y = 0.66$ in the starting materials.

TABLE I. Nominal F compositions in the starting materials ($x$) dependence on the analytical F compositions in the single crystals ($y$), $c$-axis lattice parameter and superconducting anisotropy ($\gamma_s$) in the grown single crystals.

|   |   | Nominal F compositions in the starting materials ($x$) | | |
|---|---|---|---|---|
|   |   | 0.3 | 0.7 | 0.9 |
| R : La | Analytical F compositions in the single crystals ($y$) | 0.23 | 0.46 | 0.46 |
|   | $c$-axis lattice parameter (Å) | 13.57 | 13.37 | 13.39 |
|   | Superconducting anisotropy ($\gamma_s$) | --- | 35-37 | 36-45 |
| R : Ce | Analytical F compositions in the single crystals ($y$) | 0.53 | 0.65 | 0.66 |
|   | $c$-axis lattice parameter (Å) | 13.54 | 13.39 | 13.40 |
|   | Superconducting anisotropy ($\gamma_s$) | --- | 13-21 | 14 |

| R : Nd | Analytical F compositions in the single crystals (y) | 0.26 | 0.37 | 0.38 |
|---|---|---|---|---|
| | c-axis lattice parameter (Å) | 13.56 | 13.43 | 13.41 |
| | Superconducting anisotropy ($\gamma_s$) | 30-34 | 30-31 | 37-40 |

--- : Unmeasurable at our system

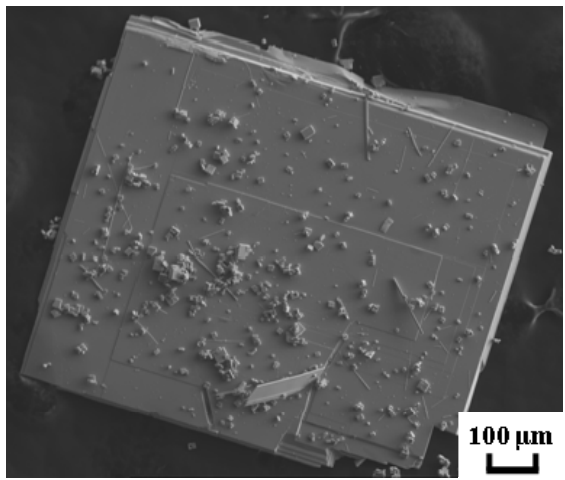

**Figure 1**

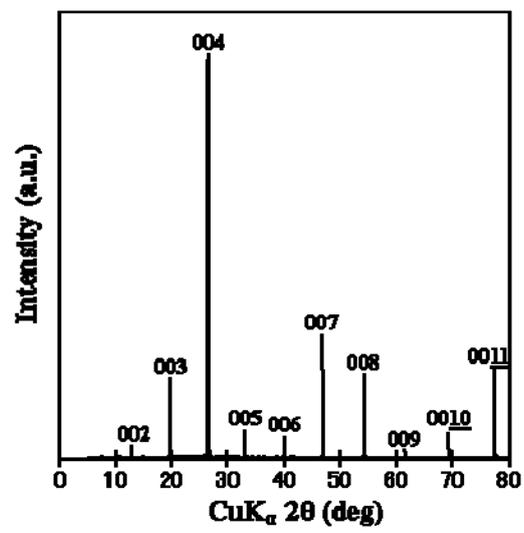

**Figure 2**

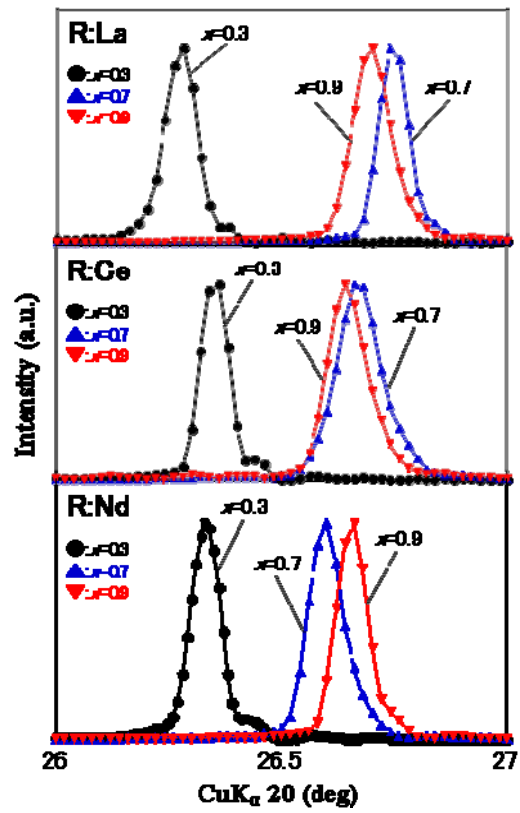

**Figure 3 (Color on the Web only)**

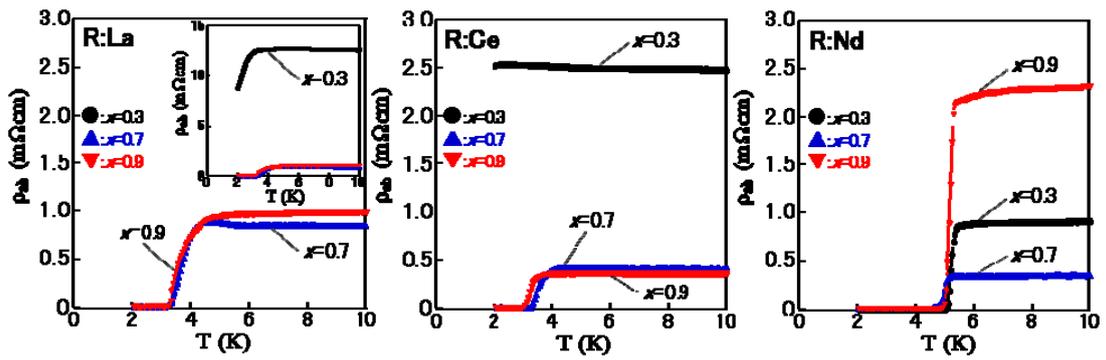

**Figure 4 (Color on the Web only)**

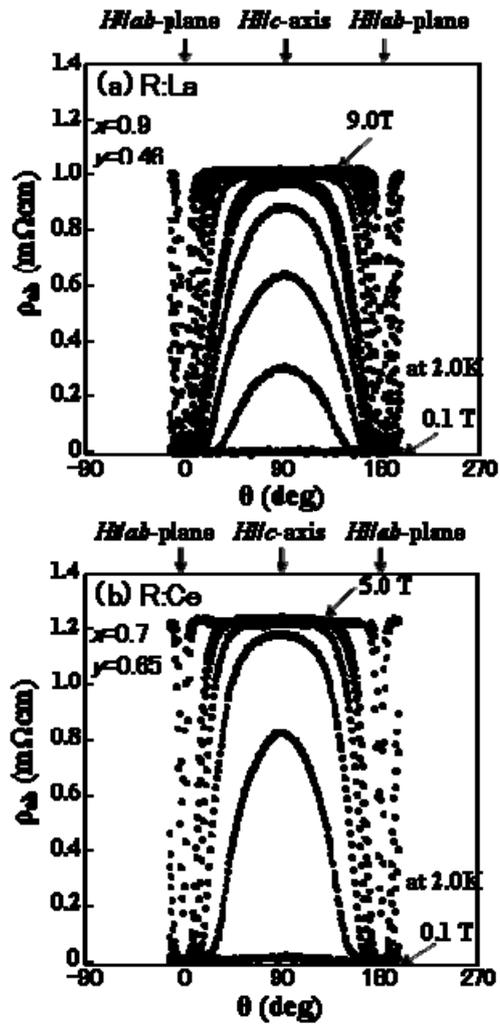

**Figure 5**

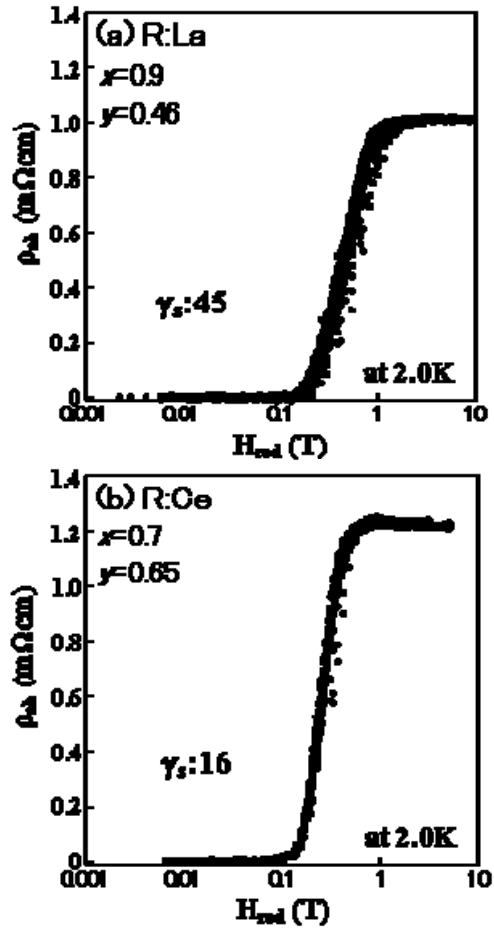

**Figure 6**

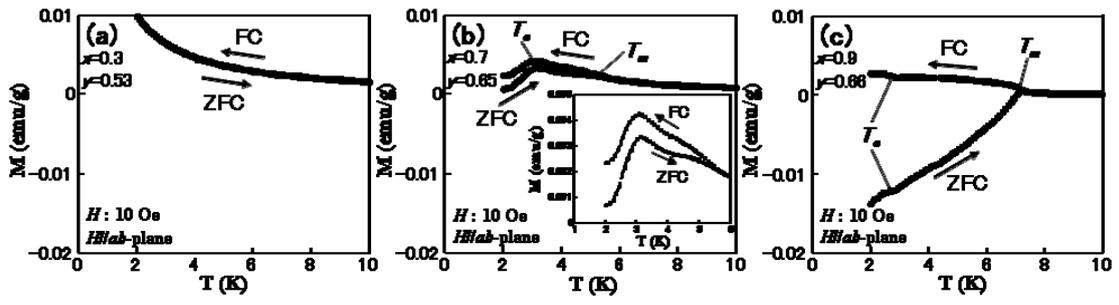

Figure 7